\documentclass[11pt]{article}
\usepackage[utf8]{inputenc}

\usepackage[a4paper, total={6in, 9in}]{geometry}
\usepackage[T1]{fontenc}
\usepackage[hyphens]{url}
\usepackage{hyperref}
\usepackage{color,soul}
\usepackage{xcolor}
\usepackage{array}
\usepackage{mhchem}
\usepackage{mathtools}
\usepackage{amsfonts}
\usepackage{amssymb}
\usepackage{cleveref}
\usepackage{fancyhdr}
\usepackage{float}
\usepackage[round]{natbib}
\usepackage{multirow}
\usepackage{makecell}
\usepackage{gensymb}
\usepackage{rotating}
\usepackage{pdflscape}
\usepackage{multirow}
\usepackage{graphicx}

\begin{document}

\title{PEACH Tree: A Multiple Sequence Alignment and Tree Display Tool for Epidemiologists}

    \begin{LARGE}

 \vspace*{6cm}
     \noindent\textbf{PEACH Tree: A Multiple Sequence Alignment and Tree Display Tool for Epidemiologists} \\
    \end{LARGE}

     \noindent\textbf{Jordan Douglas$^{1,*}$ and David Welch$^{1}$} \\
    
    \noindent $^{1}$ School of Computer Science, University of Auckland, Auckland 1010, New Zealand \\
    \noindent $^{*}$ Correspondence: jordan.douglas@auckland.ac.nz \\

\section*{Abstract}

\textbf{Summary:} PEACH Tree is an easy-to-use, online tool for displaying multiple sequence alignments and phylogenetic trees side-by-side.
PEACH Tree is powerful for rapidly tracing evolutionary and transmission histories by filtering invariant sites out of the display, and allowing samples to readily be filtered out of the display.
These features, coupled with the ability to display epidemiological metadata, make the tool suitable for  infectious disease epidemiology.
PEACH Tree further enables much needed communication between the fields of genomics and infectious disease epidemiology, as exemplified by the COVID-19 pandemic.
 \\
\textbf{Availability and Implementation:} PEACH Tree runs online at \href{https://phylopeachtree.github.io/}{https://phylopeachtree.github.io/}. \\

\clearpage
\section{Introduction}

Genomic methods are now instrumental in efforts to control infectious diseases.
While epidemiological models \citep{brauer2008compartmental} are primarily informed by case counts, trajectories, and population data -- such as movement, contact, and demography --

genomic methods exploit the fast mutation rates of certain pathogens to infer evolutionary and transmission histories \citep{grenfell04}.
Widespread pathogen sequencing forms the basis of pathogen surveillance technologies \citep{gardy2018towards} such as NextStrain \citep{hadfield2018nextstrain} and GISAID \citep{shu2017gisaid} and has informed public health response for a  range of epidemics \citep{baize2014emergence,faria2017establishment,seemann2020tracking,douglas2021phylodynamics}. 
Historically, infectious disease epidemiology and genomics have existed as distinct fields, however the rise of real-time sequencing and its ability to inform outbreak response has demonstrated the benefit in strong communication between the two.

Visualisation is a vital element in scientific communication.
Biological sequences, such as viral genomes, can be aligned and viewed using a wide range of existing programs \citep{larsson2014aliview, waterhouse2009jalview, larkin2007clustal}.
Phylogenetic trees can then be inferred from multiple sequence alignments (MSA) and displayed with a range of software packages \citep{paradis2004ape,rambaut2009figtree,vaughan2017icytree,douglas2021uglytrees}.
Viewing large MSAs or large trees are nontrivial tasks and the display can easily become overloaded with information.
However, when studying infectious diseases, the segregating sites (i.e., alignment positions which vary among the samples) are typically of more interest than the invariant sites, and some cases, outbreaks, species, or other taxonomic groups are often of more interest than others.
Displaying the full dataset may overwhelm the user with unwanted information and impede computational performance. 
For an epidemiologist in particular, the ability to easily trace transmission histories, view symptom onset dates, and link genomes to case numbers (and other metadata) are desirable features in any software package built to view infectious disease transmission.

We present PEACH Tree (\textbf{P}lotting \textbf{E}pidemiological and \textbf{A}lignment \textbf{CH}aracters onto phylogenetic \textbf{Tree}s),
a program for viewing multiple sequence alignments and phylogenetic trees specifically designed for (but not restricted to) infectious disease epidemiology and pathogen surveillance.
PEACH Tree is responsive, easy-to-use, and runs in the web-browser.

\section{PEACH Tree}

\begin{figure*}
\centering

\includegraphics[width=\textwidth]{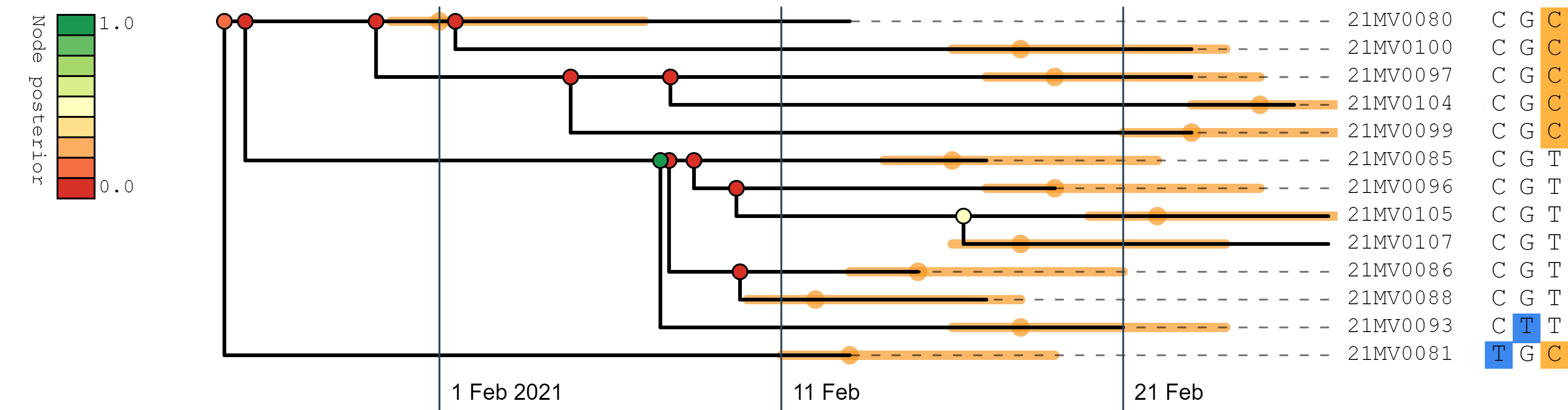}
\includegraphics[width=\textwidth]{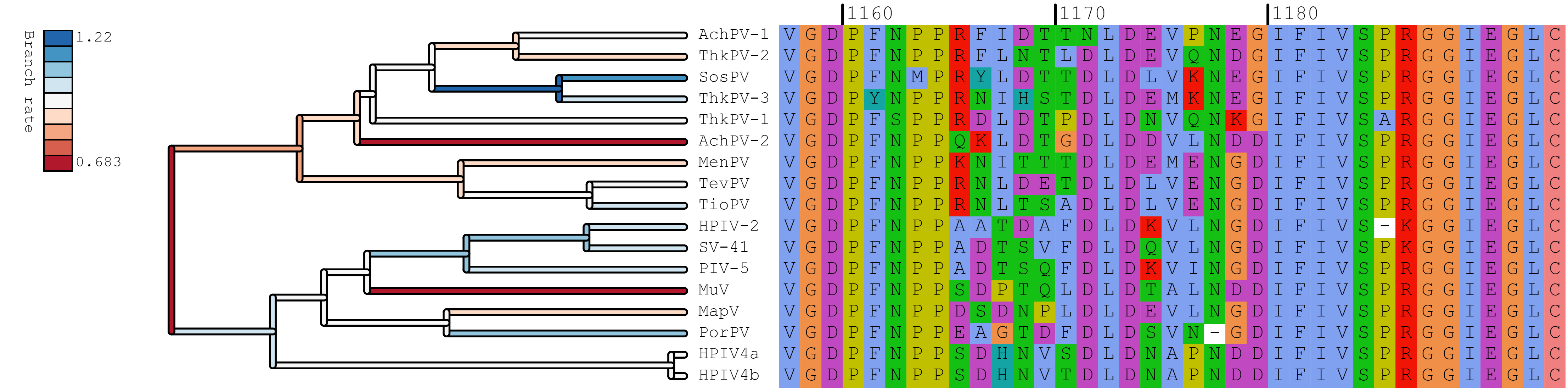}
\vspace*{-.3in}
\caption{Two PEACH Tree outputs showing trees (left) constructed from multiple sequence alignments (right).
Top: transmission tree for the New Zealand February 2021 COVID-19 outbreak \citep{douglas2021real}.
Only 13 [monophyletic] cases from the full tree are displayed along with their segregating sites.
Sites are coloured by minor alleles. 
Internal nodes are coloured by their posterior clade support probability and the infectious periods are displayed as orange bars (modelled by symptom onset dates minus 2 days, plus 5 days). 
Note that onset dates were randomised for privacy protection.
Bottom: phylogenetic tree of \emph{Rubulavirinae} L proteins \citep{douglas2021evolutionary}, with branches coloured by their substitution rate under the relaxed clock model \citep{douglas2021adaptive}.
All sites (within the indicated range) are displayed.}
\label{fig:fig1}
\end{figure*}

When opening PEACH Tree, the user is prompted for an MSA (FASTA format) and/or a tree (Newick/NEXUS format). 
If a tree is not provided, a neighbour joining tree can be constructed from the MSA \citep{saitou1987neighbor}.
The user may also upload case metadata (comma- or tab-separated-variable format), describing sample dates or symptom onset dates for instance.
PEACH Tree then plots the phylogenetic tree alongside the MSA and renders further epidemiological annotations onto the display (\textbf{Figure \ref{fig:fig1}}).
By default, only segregating sites are shown, as opposed to the complete MSA.
A subset of samples (such as a monophyletic group) can be focused on and the segregating sites are recalculated.
This can be useful for understanding genomic variants within a particular outbreak or cluster, for instance.
The tree can be displayed as a transmission tree, where internal nodes are oriented to represent a transmission event from the top child to the bottom child. 
The orientation of nodes can be flipped by clicking on the transmission node.
A Scalable Vector Graphics (SVG) or Portable Network Graphics (PNG) file can be readily downloaded from the web-browser. 






While numerous MSA and tree visualisation tools already exist, PEACH Tree stands alone for several reasons.
First, PEACH Tree is a fast and responsive web-browser implementation. 
Second, PEACH Tree displays MSAs and trees together side-by-side.
Third, PEACH Tree is specifically catered to the domain of infectious disease transmission and comes with a range of tools for easily focusing on samples and sites, and displaying epidemiological annotations.

\section{Funding}

This project was funded by the New Zealand Ministry of Health and  Ministry of Business, Innovation and Employment.

\clearpage

\begin{thebibliography}{20}
\providecommand{\natexlab}[1]{#1}
\providecommand{\url}[1]{\texttt{#1}}
\expandafter\ifx\csname urlstyle\endcsname\relax
  \providecommand{\doi}[1]{doi: #1}\else
  \providecommand{\doi}{doi: \begingroup \urlstyle{rm}\Url}\fi

\bibitem[Baize et~al.(2014)Baize, Pannetier, Oestereich, Rieger, Koivogui,
  Magassouba, Soropogui, Sow, Ke{\"\i}ta, De~Clerck,
  et~al.]{baize2014emergence}
Sylvain Baize, Delphine Pannetier, Lisa Oestereich, Toni Rieger, Lamine
  Koivogui, N'Faly Magassouba, Barr{\`e} Soropogui, Mamadou~Saliou Sow, Sakoba
  Ke{\"\i}ta, Hilde De~Clerck, et~al.
\newblock Emergence of zaire ebola virus disease in guinea.
\newblock \emph{New England Journal of Medicine}, 371\penalty0 (15):\penalty0
  1418--1425, 2014.

\bibitem[Brauer(2008)]{brauer2008compartmental}
Fred Brauer.
\newblock Compartmental models in epidemiology.
\newblock In \emph{Mathematical epidemiology}, pages 19--79. Springer, 2008.

\bibitem[Douglas(2021)]{douglas2021uglytrees}
Jordan Douglas.
\newblock Uglytrees: a browser-based multispecies coalescent tree visualizer.
\newblock \emph{Bioinformatics}, 37\penalty0 (2):\penalty0 268--269, 2021.

\bibitem[Douglas et~al.(2021{\natexlab{a}})Douglas, Drummond, and
  Kingston]{douglas2021evolutionary}
Jordan Douglas, Alexei~J Drummond, and Richard~L Kingston.
\newblock Evolutionary history of cotranscriptional editing in the
  paramyxoviral phosphoprotein gene.
\newblock \emph{Virus Evolution}, 7\penalty0 (1):\penalty0 veab028,
  2021{\natexlab{a}}.

\bibitem[Douglas et~al.(2021{\natexlab{b}})Douglas, Geoghegan, Hadfield,
  Bouckaert, Storey, Ren, de~Ligt, French, and Welch]{douglas2021real}
Jordan Douglas, Jemma~L Geoghegan, James Hadfield, Remco Bouckaert, Matthew
  Storey, Xiaoyun Ren, Joep de~Ligt, Nigel French, and David Welch.
\newblock Real-time genomics for tracking severe acute respiratory syndrome
  coronavirus 2 border incursions after virus elimination, new zealand.
\newblock \emph{Emerging Infectious Diseases}, 27\penalty0 (9):\penalty0 2361,
  2021{\natexlab{b}}.

\bibitem[Douglas et~al.(2021{\natexlab{c}})Douglas, Mendes, Bouckaert, Xie,
  Jim{\'e}nez-Silva, Swanepoel, de~Ligt, Ren, Storey, Hadfield,
  et~al.]{douglas2021phylodynamics}
Jordan Douglas, F{\'a}bio~K Mendes, Remco Bouckaert, Dong Xie, Cinthy~L
  Jim{\'e}nez-Silva, Christiaan Swanepoel, Joep de~Ligt, Xiaoyun Ren, Matt
  Storey, James Hadfield, et~al.
\newblock Phylodynamics reveals the role of human travel and contact tracing in
  controlling the first wave of covid-19 in four island nations.
\newblock \emph{Virus Evolution}, 7\penalty0 (2):\penalty0 veab052,
  2021{\natexlab{c}}.

\bibitem[Douglas et~al.(2021{\natexlab{d}})Douglas, Zhang, and
  Bouckaert]{douglas2021adaptive}
Jordan Douglas, Rong Zhang, and Remco Bouckaert.
\newblock Adaptive dating and fast proposals: Revisiting the phylogenetic
  relaxed clock model.
\newblock \emph{PLoS computational biology}, 17\penalty0 (2):\penalty0
  e1008322, 2021{\natexlab{d}}.

\bibitem[Faria et~al.(2017)Faria, Quick, Claro, Theze, de~Jesus, Giovanetti,
  Kraemer, Hill, Black, da~Costa, et~al.]{faria2017establishment}
Nuno~R Faria, Joshua Quick, IM~Claro, Julien Theze, Jacqueline~G de~Jesus,
  Marta Giovanetti, Moritz~UG Kraemer, Sarah~C Hill, Allison Black, Antonio~C
  da~Costa, et~al.
\newblock Establishment and cryptic transmission of zika virus in brazil and
  the americas.
\newblock \emph{Nature}, 546\penalty0 (7658):\penalty0 406--410, 2017.

\bibitem[Gardy and Loman(2018)]{gardy2018towards}
Jennifer~L Gardy and Nicholas~J Loman.
\newblock Towards a genomics-informed, real-time, global pathogen surveillance
  system.
\newblock \emph{Nature Reviews Genetics}, 19\penalty0 (1):\penalty0 9--20,
  2018.

\bibitem[Grenfell et~al.(2004)Grenfell, Pybus, Gog, Wood, Daly, Mumford, and
  Holmes]{grenfell04}
Bryan~T Grenfell, Oliver~G Pybus, Julia~R Gog, James L~N Wood, Janet~M Daly,
  Jenny~A Mumford, and Edward~C Holmes.
\newblock Unifying the epidemiological and evolutionary dynamics of pathogens.
\newblock \emph{Science}, 303:\penalty0 327--332, 2004.

\bibitem[Hadfield et~al.(2018)Hadfield, Megill, Bell, Huddleston, Potter,
  Callender, Sagulenko, Bedford, and Neher]{hadfield2018nextstrain}
James Hadfield, Colin Megill, Sidney~M Bell, John Huddleston, Barney Potter,
  Charlton Callender, Pavel Sagulenko, Trevor Bedford, and Richard~A Neher.
\newblock Nextstrain: real-time tracking of pathogen evolution.
\newblock \emph{Bioinformatics}, 34\penalty0 (23):\penalty0 4121--4123, 2018.

\bibitem[Larkin et~al.(2007)Larkin, Blackshields, Brown, Chenna, McGettigan,
  McWilliam, Valentin, Wallace, Wilm, Lopez, et~al.]{larkin2007clustal}
Mark~A Larkin, Gordon Blackshields, Nigel~P Brown, R~Chenna, Paul~A McGettigan,
  Hamish McWilliam, Franck Valentin, Iain~M Wallace, Andreas Wilm, Rodrigo
  Lopez, et~al.
\newblock Clustal w and clustal x version 2.0.
\newblock \emph{bioinformatics}, 23\penalty0 (21):\penalty0 2947--2948, 2007.

\bibitem[Larsson(2014)]{larsson2014aliview}
Anders Larsson.
\newblock Aliview: a fast and lightweight alignment viewer and editor for large
  datasets.
\newblock \emph{Bioinformatics}, 30\penalty0 (22):\penalty0 3276--3278, 2014.

\bibitem[Paradis et~al.(2004)Paradis, Claude, and Strimmer]{paradis2004ape}
Emmanuel Paradis, Julien Claude, and Korbinian Strimmer.
\newblock Ape: analyses of phylogenetics and evolution in r language.
\newblock \emph{Bioinformatics}, 20\penalty0 (2):\penalty0 289--290, 2004.

\bibitem[Rambaut(2009)]{rambaut2009figtree}
A~Rambaut.
\newblock Figtree v1. 3.1.
\newblock \emph{http://tree. bio. ed. ac. uk/software/figtree/}, 2009.

\bibitem[Saitou and Nei(1987)]{saitou1987neighbor}
Naruya Saitou and Masatoshi Nei.
\newblock The neighbor-joining method: a new method for reconstructing
  phylogenetic trees.
\newblock \emph{Molecular biology and evolution}, 4\penalty0 (4):\penalty0
  406--425, 1987.

\bibitem[Seemann et~al.(2020)Seemann, Lane, Sherry, Duchene, da~Silva, Caly,
  Sait, Ballard, Horan, Schultz, et~al.]{seemann2020tracking}
Torsten Seemann, Courtney~R Lane, Norelle~L Sherry, Sebastian Duchene,
  Anders~Gon{\c{c}}alves da~Silva, Leon Caly, Michelle Sait, Susan~A Ballard,
  Kristy Horan, Mark~B Schultz, et~al.
\newblock Tracking the covid-19 pandemic in australia using genomics.
\newblock \emph{Nature communications}, 11\penalty0 (1):\penalty0 1--9, 2020.

\bibitem[Shu and McCauley(2017)]{shu2017gisaid}
Yuelong Shu and John McCauley.
\newblock Gisaid: Global initiative on sharing all influenza data--from vision
  to reality.
\newblock \emph{Eurosurveillance}, 22\penalty0 (13):\penalty0 30494, 2017.

\bibitem[Vaughan(2017)]{vaughan2017icytree}
Timothy~G Vaughan.
\newblock Icytree: rapid browser-based visualization for phylogenetic trees and
  networks.
\newblock \emph{Bioinformatics}, 33\penalty0 (15):\penalty0 2392--2394, 2017.

\bibitem[Waterhouse et~al.(2009)Waterhouse, Procter, Martin, Clamp, and
  Barton]{waterhouse2009jalview}
Andrew~M Waterhouse, James~B Procter, David~MA Martin, Mich{\`e}le Clamp, and
  Geoffrey~J Barton.
\newblock Jalview version 2—a multiple sequence alignment editor and analysis
  workbench.
\newblock \emph{Bioinformatics}, 25\penalty0 (9):\penalty0 1189--1191, 2009.

\end{thebibliography}

\end{document}